\documentclass[a4paper,12pt]{article}
\usepackage{amsmath}
\usepackage{pstricks}
\usepackage{pst-node}
\usepackage[ansinew]{inputenc}
\usepackage{amssymb,amsmath}
\usepackage{amsfonts}
\usepackage{epsfig}
\usepackage[english]{babel}
\usepackage{graphics}

\def\a{\alpha}
\def\b{\beta}
\def\g{\gamma}

\def\e{\varepsilon}

\font\Sets=msbm10

\def\Integer {\hbox{\Sets Z}}    \def\Real {\hbox{\Sets R}}

   \def\Natural {\hbox{\Sets N}}

 \def\Rational {\hbox{\Sets Q}}

\def\be{\begin{equation}}       \def\ba{\begin{array}}

\def\ee{\end{equation}}         \def\ea{\end{array}}

\def\bea {\begin{eqnarray}}      \def\eea {\end{eqnarray}}

\def\bean{\begin{eqnarray*}}    \def\eean{\end{eqnarray*}}

\def\eps{\varepsilon}

\def\<{\langle} \def\({\left(}  \def\>{\rangle} \def\){\right)}

\newtheorem{exi}{Example}

\author{Elena Kartashova\footnote{Author acknowledges support of
the Austrian Science Foundation (FWF) under projects SFB F013/F1304.}\\
RISC, J. Kepler University,\\ Altenbergerstr. 69, 4040 Linz, Austria\\
e-mail: lena@risc.uni-linz.ac.at}

\title{A model of laminated wave turbulence}

\begin{document}
%generates titel
\date{}
\maketitle

%\tableofcontents {

\abstract{A model of laminated wave turbulence  is presented. This
model consists of two co-existing layers - one with continuous
waves' spectra, covered by KAM theory and Kolmogorov-like power
spectra, and one with discrete waves' spectra, covered by discrete
classes of waves and Clipping method. Some known laboratory
experiments and numerical simulations are explained in the frame
of this model. }\\


{\bf 1. Continuous wave spectra}

In \cite{Kol1} Kolmogorov presented energy spectrum of turbulence
 describing the distribution of the energy among turbulence
vortices as function of vortex size and thus founded the field of
mathematical analysis of turbulence. Kolmogorov regarded some
inertial range of wave numbers, between viscosity and dissipation,
and suggested that at this range, turbulence is (1) locally
homogeneous (no dependence on position) and (2) locally isotropic
(no dependence on direction) which can be summarized as follows:
probability distribution for the relative velocities of two
particles in the fluid only depends on the distance between
particles. Using these suggestions and dimensional analysis,
Kolmogorov deduced that energy distribution, called now Kolmogorov´s
spectrum, is proportional to $k^{-5/3}$ for wave numbers $k$.
Results of numerical simulations and real experiments carried out to
prove this theory are somewhat contradictious. On the one hand,
probably the most spectacular example of the validity of
Kolmogorov´s spectra is provided in \cite{Grant} where measurements
in tidal currents near Seymour Narrows north of Campbell River on
Vancouver Island were described and $-5/3$ spectra appeared at the
range of $10^4$ (energy dissipation at a scale of  millimeters and
energy input - at 100 meters). On the other hand, Kolmogorov´s
spectra have been obtained under the assumptions opposite to
Kolmogorov´s \cite{Chor} so that exponent $-5/3$ corresponds to both
direct and inverse cascades.

With a hope to diminish established unclearness of Kolmogorov´s
theory in a more simple setting, theory of wave (or weak) turbulence
(WT) in the systems with continuous wave spectra has been developed.
The problem is regarded in the very general form \be
\label{wt}L(\psi)=-\eps N(\psi) \ee where $L$ and $N$ are linear and
nonlinear operators consequently, $0<\eps << 1$ is a parameter of
nonlinearity and linear part possesses wave-like solutions of the
form
 \be \label{linwave}\psi
(\vec{x})=A(\vec{k}, x)\exp{i(\vec{k}\vec{x} - \omega (\vec{k})
t)}.\ee
 Choice of $\eps$ is
defined by specifics of the physical wave system under the study.
For instance, for spherical planetary waves $\eps$ is usually chosen
as the ratio of the particle velocity  to the phase velocity while
for water waves usually $\eps = a |\vec k|$ , where $a$ is amplitude
of a wave and $\eps$ characterizes in this case the steepness of the
waves. For small enough $\eps$, solutions of Eq.(\ref{wt}) are
described at the slow-time scales $T_1=t/\eps, \ T_2=t/\eps^2, \
T_3=t/\eps^3, ..., $ etc. by resonantly interacting waves only, i.e.
by the waves with wave-vectors satisfying to resonant conditions
(for nonlinearity of order $n$):
\begin{eqnarray}\label{res}
\begin{cases}
\omega (\vec k_1) \pm \omega (\vec k_2)\pm ... \pm \omega (\vec k_{n+1}) = 0,\\
\vec k_1 \pm \vec k_2 \pm ... \pm \vec k_{n+1} = 0
\end{cases}
\end{eqnarray}
 so that quadratic nonlinearity corresponds to
3-waves interactions, cubic - to 4-waves interactions, etc.
Obviously, in every physical problem the resonances have some
nonzero width, i.e. Eq.(\ref{res}.1) takes form \be \label{Delta}
\omega (\vec k_1) \pm \omega (\vec k_2)\pm ... \pm \omega (\vec
k_{n+1}) = \Delta \ee with a small but nonzero discrepancy
$0<\Delta<<1$. Taking this into account,
 nonlinear part of (\ref{wt})
can be rewritten as
 \be\label{sigma}
\Sigma_{i}  \frac{V_i \delta (\vec k_1 \pm \vec k_2 \pm ... \pm \vec
k_{i})} {\omega (\vec k_1) \pm \omega (\vec k_2)\pm ... \pm \omega
(\vec k_{i}) } \ee where $\delta$ is a delta-function and each term
of this sum with nonzero vertex coefficient $V_i$ corresponds to a
specific slow-time scale of wave interactions. The representation
(\ref{sigma}) is used then for construction of a wave kinetic
equation, with corresponding vertex coefficients and delta-functions
in the under-integral expression.

From mathematical point of view, it is important to establish the
{\bf finiteness} of nonlinearity given by (\ref{sigma}) in the case
when $\Delta \rightarrow 0$ because representation (\ref{sigma})
becomes meaningless if $\Delta=0$.
 This problem - so-called "problem of small
denominators" - was solved by KAM theory
(\cite{Kol3},\cite{Arn1},\cite{Mos}) in the following way. For small
enough $\Delta$
 and
{\bf sufficiently} irrational dispersion function $\omega$, these
wave systems contain an infinite set of invariant tori which carry
quasi-periodic motions which in phase space are confined to the
tori. Main result of KAM theory is therefore a decomposition of
action into disjoint invariant sets, and though it contradicts
ergodicity but not very substantially as the size of the system
tends to infinity \cite{Arn2}. In particularly, random phase
approximation can be assumed and kinetic equations and
Kolmogorov-like power spectra $k^{\g}, \ \g<0, $ give then
appropriate description of these wave systems at the corresponding
time scales. It means that some special set of points in spectral
space, corresponding to the $\Delta$-vicinity of exact resonances,
has been excluded from consideration in order to obtain  wave
kinetic equation.
We give detailed description of this subset of spectral space in the section 4 of this paper. \\

{\bf 2. Discrete wave spectra}

There exist a lot of wave phenomena which are due to discreteness of
the wave spectra (corresponds to zero- or periodic boundary
conditions)
 and  can not be explained
in terms of kinetic equations and power energy spectra. To describe
these phenomena, WT in the systems with discrete spectra has been
developed \cite{PRL},\cite{AMS}. It turned out that discrete systems
possess some qualitatively new properties: (a) all resonantly
interacting waves are divided into disjoint classes, there is no
energy flow between different these classes; (b) major part of the
waves do not interact; (c) all interactions of a specific wave are
confined to some finite domain; (d) number of interacting waves
depends on the form of boundary conditions, for the great number of
boundary conditions interactions are not possible; (e) all
properties (a)-(d) keep true for approximate interactions, i.e. for
some
 small enough discrepancy $0<\Delta<<1$. This fact gave a rise
to Clipping method \cite{Clip} which allows "to clip out" all
non-interacting waves from the whole spectra and study only those
which do interact, exactly or approximately. Approximate resonances
are understood on a discrete lattice, i.e. wave vectors of
approximately interacting waves are also integers. The energetic
behavior of these  systems is described then independently (at each
slow-time scale $T_j$) by a few small
 systems of ordinary differential equations (SODE) on
slowly-changing amplitudes of resonantly interacting waves, i.e.
amplitude of linear wave (\ref{linwave}) is a function of some $T_j$
depending on the form on nonlinearity $N$. For instance, for 3-waves
interactions major part of these SODE consist of three equations on
three (real-valued) amplitudes and can be solved explicitly in terms
of elliptic functions on $T_1$. To compare with WT of the systems
with continuous spectra, SODE are to be used instead of kinetic
equation and their coupling coefficients - instead of power-law
spectra. Notice that though kinetic description does not apply for
discrete systems, some of these results are in a sense similar to
those of KAM theory, for instance, Theorem on the partition
\cite{PHD1} can be regarded as
an analog of KAM-Theorem for discrete systems.\\

{\bf 3. Transition from discrete to continuous spectra}

 Now, the standard qualitative model of
the WT can be presented as follows:  short waves are described by
Kolmogorov´s energy spectra and kinetic equations, long waves are
described by Clipping method and dynamic equations, and somewhere in
between a "transition" interval exists that has its own specifics
and should be described separately.

We would like to demonstrate  some contradictiveness of this
qualitative model and begin with two remarks. (1) WT of discrete
waves systems has been developed for arbitrary wave numbers which
means that transition from finite to infinite domain can be
constructed not only in some finite "transition" interval but at the
whole infinite range of wave numbers. (2) Transition from discrete
to continuous spectrum
 is often regarded in somewhat over-simplified way: if say, real-valued wave vectors
$\vec{k}=(k_x,k_y)$ have dispersion function $\omega(k_x,k_y)$ with
$ k_x,k_y \in \Real$, then the same function of integer variables,
$\omega(m,n)$ with $ m,n \in \Integer$, describes corresponding
discrete waves. In general, it is not true. We demonstrate it taking
barotropic vorticity equation (BVE),  also known as
Obukhov-Charney-Hasegawa-Mima equation, as our main example
motivated by its wide applicability for describing a great number of
physically important phenomena in astrophysics, geophysics and
plasma physics.

BVE on a sphere has form \be\label{BVE1} \frac{\partial \triangle
\psi}{\partial t} + 2 \frac{\partial \psi}{\partial \lambda} +
J(\psi,\triangle  \psi) =0 \ee with  linear waves of the form  \be
\psi_{sphere} = A P_n^m (\sin \phi) \exp{i[m \lambda
+\omega_{sphere} t]} .\ee Here $\psi$ is the stream-function;
variables $t, \phi$ and $\lambda$ physically mean the time, the
latitude ($-\pi/2 \leq \phi \leq \pi/2$) and the longitude ($0 \leq
\lambda \leq 2\pi$) respectively; $P_n^m (x)$ is the associated
Legendre function of degree $n$ and order $m$. The same equation
taken on infinite $\b$-plane has linear waves of the form
$$
\psi_{plane} = A \exp {i(k_{x}x + k_{y}y+\omega_{plane}t)},
$$
which means that
$$
\omega_{sphere} = m /[n(n+1)] \quad \mbox{and} \quad \omega_{plane}
= k_{x}/(1+k^2_{x}+k^2_{y}),
$$
here constant multipliers are omitted because they disappear due to
homogeneous form of Eq.(\ref{res}.1). It is easy to see that no wave
vectors $\vec{k}=(m,n):m,n \in \Integer$ satisfy Eq.(\ref{res}.1)
with $\omega_{sphere}$ and with $\omega_{plane}$ simultaneously. It
means that discrete waves do not have images on infinite plane when
such a "naive" transition is regarded.

More intrinsic construction of the transition from spherical to
plane planetary waves \cite{KPR} can be derived in following way.
Regarding $m \sim n >> 1$ and using asymptotic approximation for
Legendre functions, one can "convert" (not always but in a bounded
latitudinal belt with the width $\sim n^{-1}$) one spherical wave
into a linear combination of two plane waves
$$
A \exp {i(k(\varphi_0)_{x}x \pm k(\varphi_0)_{y}y+\omega_{plane}t)},
$$
where local wave numbers $k(\varphi_0)_{x}, k(\varphi_0)_{y} \in
\Real$ are functions of the initial spherical wave number $m,n$ and
of the so-called interaction latitude $\varphi_0$:
$$
\cos^2\varphi_0=\frac{m_1^2(n_2^2+n_3^2-n_1^2)+
m_2^2(n_1^2+n_3^2-n_2^2)+m_3^2(n_1^2+n_2^2-n_3^2))}{n_1^2n_2^2
+n_1^2n_3^2+n_2^2n_3^2 -(n_1^4+n_2^4+n_3^4)/4}.
$$
If interaction latitude exists, $0<\cos^2\varphi_0<1$, plane images
of spherical waves interact as in classical $\b$-plane
approximation. In particularly, this means that (1) transition from
a spherical domain to an infinite plane is  transition to a
one-parametric family of infinite planes, and (2) such a transition
is not always possible. A very important fact is that plane wave
system keeps memory about spherical interactions: coupling
coefficient of the plane images of spherical waves is $\sim k^{3/2}$
and $\sim k^{7/6}$ otherwise and $k=|\vec{k}|$.

The same reasoning allows to construct a transition from a square
domain (dispersion function being then
$\omega_{square}=1/\sqrt{m^2+n^2}$) to the infinite $\b$-plane where
difference in magnitudes of coupling coefficients is even more
substantial: $\sim k^2$ for plane images of the waves from square
domain and $\sim k$ otherwise \cite{KR}. These results hold for
discrete approximate interactions in following way: long-wave part
of spectrum is dominated by a few resonantly interacting waves with
huge amplitudes while short-wave part of the spectrum consists of
many approximately interacting waves with substantially smaller
amplitudes. But in any case, coupling coefficients are of order
$k^{\g}$ with $\g>0$ which {\bf apparently} contradicts to the
existence of Kolmogorov-like power spectra $k^{\g}$ with
$\g<0$ in the region of short waves.\\

{\bf 4. Laminated WT}

In order to resolve this apparent contradiction we have go back to
the very base of the KAM theory. Its main results are based on the
famous Thue theorem, giving low estimate for the distance between
any algebraic number $\a$ of degree $n>2$ and a rational number $p/q
\in \Rational$:
$$|\a-\frac{p}{q}|> \frac{c(\a)}{q^{\Delta+1+n/2}}, \quad \forall
\e>0$$ where $c(\a)$ is a constant depending on $\a$ and $\Delta$
can be arbitrary small. This fact allows to construct KAM tori, with
$\a$
 being a ratio of frequencies of interacting
waves, and KAM theorem states then that {\bf almost} all tori are
preserved. "Almost" means in particular that tori with rationally
related frequencies  (corresponds to $\a$ being an algebraic number
of degree $n=1$) are explicitly {\bf excluded} from consideration.
Since the union of invariant tori has positive Liouville measure and
$\Rational$ has measure 0, this exclusion is supposed to be not very
important.

Coming back to the example of spherical planetary waves, one can see
immediately that the ratio of the frequencies for resonantly
interacting waves in this case is a rational number, both for exact
and approximate interactions. Therefore, these waves are not
described by KAM theory. Obviously, the same keeps true for an
arbitrary wave system with rational dispersion function $\omega.$
Case of planetary waves in a square domain is a bit more
complicated. It is proven \cite{AMS} that exact resonances in this
case are described by the wave vectors $\vec{k}_i=(m_i,n_i)$
satisfying the (necessary) condition
$$
k_i=a_i\sqrt{q}, \ a_i \in  \Natural \  \ \forall i=1,2,3,
$$
with {\bf the same} square-free $q$. This fact allows to construct
disjoint classes of resonantly interacting waves and $q$ is called
index of the class. Obviously, for the waves belonging to the same
class, the ratio of their frequencies is a rational number,
$\omega_i/\omega_j \in \Rational , \ \forall i,j$ and these waves
are excluded from KAM theory. Waves, interacting approximately, may
have different indices (not necessarily) but the ratio of
frequencies $\omega_i/\omega_j$ is then an algebraic number of
degree $\leq 2$ due to the form of dispersion function and,
therefore, these waves are also excluded from the KAM theory.
Similar results can be proven for exact resonances in many wave
systems, for instance, for an arbitrary wave system in which
dispersion function is a polynomial of finite degree on $k$  with at
least one non-zero coefficient in front of an odd degree of $k$.

Let us summarize the results obtained. Continuous WT (CWT) describes
energetic behavior of a wave system for the short-waves' part of
spectrum {\bf excluding} nodes of rational lattice thus leaving some
gaps in the spectrum which are supposed to be not important in
short-waves' part. Discrete WT (DWT) fills these gaps all over the
spectrum. In fact we have two layers of turbulence - CWT (layer I)
and DWT (layer II), which are mutually complementary and should be
regarded simultaneously.

Layer I provides KAM tori and stochastic enough turbulence in
short-waves range with Kolmogorov´s spectra in the inertial
interval; direct or inverse energy cascades are possible;
wave-numbers range of energy pumping influences the results.

Layer II provides a countable number of waves with big amplitudes
all over the wave spectrum; some of the waves do not change their
energies (non-interacting waves) and others do exchange energy
within small independent groups; there is no energy cascade at this
layer; results do not depend on the  wave-numbers range of energy
pumping.

The co-existence of these two layers means in particularly that in
the waves' range, classically described by Kolmogorov-like spectra,
there exist also a small group of waves from layer II with
substantially bigger energies than their "neighbors" from layer I,
and this small group generates appearance of some structures. We
give here a few examples of known phenomena which can be explained
in the same frame of the model of laminated turbulence.

(1) Very clear example of the co-existence of these two layers is
given in \cite{zak3} where turbulence of capillary waves was studied
in the frame of simplified dynamical equations for the potential
flow of an ideal incompressible fluid. A stationary regime of
so-called "frozen turbulence" had been discovered: in small
wave-numbers region wave spectrum consists of "several dozens of
excited low-number harmonics" which construct "ring structures in
the spectrum of surface elevation". The appearance of these
structures does not depend on the damping and pumping, and in all
computations "the Kolmogorov´s spectrum coexists with the spectrum
of another, "frozen" type, concentrated in the region of low
wave-numbers and fastly decreasing to large wave-numbers. If the
level of nonlinearity is low enough, such "frozen" regimes are
dominant" (\cite{zak3}). Obviously, these ring structures are due to
non-interacting waves of layer II and similar structures were also
observed in laboratory experiments and identified as such
\cite{ham}. Notice that as there exists no exact three-wave
interactions among capillary waves with $\omega^2=k^3$ \cite{AMS},
we observe in regime of frozen turbulence of capillary waves only
discrete waves with constant amplitudes.

(2) Similar experiments/simulations with, say, four-waves
interactions among gravity waves, $\omega^2=k$, will demonstrate
that frozen turbulence partly "thaws out"  because also changes in
the amplitudes of resonantly interacting discrete waves should be
observed (cf. "bursty" spectrum in \cite{naz}).

(3) Mesoscopic turbulence \cite{zak4} (corresponds to "transition"
interval mentioned above) discovered in numerical experiments on
modelling of turbulence of gravity waves on the surface of deep
ideal incompressible fluid gives another example of manifestation of
laminated turbulence. It was established
 that not all waves "have the same rights" and the existence of
 so-called "elite society of harmonics"
 has been demonstrated; their number  amounts to only $6\%$ of the total
 number of harmonics being $10^4$  but they play the most
 active role in mesoscopic turbulence. These elite harmonics
 correspond to exact and approximate resonances of the discrete waves (layer
 II). The number of these
 harmonics depends, of course, on the specific wave system. For
 instance computations with spherical planetary waves in the
 domain of wave numbers $0<m,n \leq 1000$ (which is far beyond the region of applicability
 of BVE) shows that  total number of harmonics taking part in exact
 resonant interactions is
 22683, i.e. slightly more than $2\%.$ Consideration of
 approximate interactions of the layer II increases this amount
 but not substantially.

 (4) Zonal extended vertices (flows in
latitudinal direction) in the atmosphere can possibly be explained
in terms of plane images of spherical waves with coupling
coefficient $n^{3/2}$.

At the end of this letter we would like to make one important
remark.  To describe the short waves of layer II it is necessary to
develop fast algorithms of solving Diophantine equations in very big
integers (of the order $10^{12}$ and more). We consider it possible
basing on the existence of disjoint classes of waves participating
in a single solution. This is our current object of interest.

 Author is sincerely grateful to V.E. Zakharov for
his encouragement and permanent interest for this work, and whose
numerous results in the theory of wave turbulence gave us necessary
insights while creating a model of laminated turbulence. Author is
also very much obliged to anonymous referees for their valuable
remarks and suggestions.

\end{document}